\documentclass[aip,reprint,showpacs]{revtex4-1}
\usepackage{graphicx}
\graphicspath{{pict/}{}}
\usepackage{bm}%
\usepackage{dcolumn}% Align table columns on decimal point

\newcounter{Fig}

\newcommand{\be}{\begin{equation}}
\newcommand{\ee}{\end{equation}}

\begin{document}
%\makeatletter
 %   \def\@cite#1#2{\textsuperscript{{#1\if@tempswa #2\fi}}}
%\makeatother
\title{Complete spectral gap in coupled dielectric waveguides embedded into metal}
\author{Wei Liu}
\email{wli124@physics.anu.edu.au}
\affiliation{Nonlinear Physics
Center, Research School of Physics and Engineering, Australian
National University, Canberra, ACT 0200, Australia}
\author{Andrey Sukhorukov}
\affiliation {Nonlinear Physics Center, Research School of Physics
and Engineering, Australian National University, Canberra, ACT 0200,
Australia}
\author{Andrey Miroshnichenko}
\affiliation {Nonlinear Physics Center, Research School of Physics
and Engineering, Australian National University, Canberra, ACT 0200,
Australia}
\author {Chris Poulton}
\address{Department of Mathematical Sciences, University of
Technology, Sydney, NSW 2007, Australia}
\author {Zhiyong Xu}
\affiliation {Nonlinear Physics Center, Research School of Physics
and Engineering, Australian National University, Canberra, ACT 0200,
Australia}
\author {Dragomir Neshev}
\affiliation {Nonlinear Physics Center, Research School of Physics
and Engineering, Australian National University, Canberra, ACT 0200,
Australia}
\author {Yuri Kivshar}
\affiliation {Nonlinear Physics Center, Research School of Physics
and Engineering, Australian National University, Canberra, ACT 0200,
Australia}
\begin{abstract}
We study a  plasmonic coupler involving backward (TM$_{01}$) and
forward (HE$_{11}$) modes of dielectric waveguides embedded into
infinite metal. The simultaneously achievable contradirectional
energy flows and codirectional wavevectors in different channels
lead to a spectral gap, despite the absence of periodic structures
along the waveguide. We demonstrate that a complete spectral gap can
be achieved in a symmetric structure composed of four coupled
waveguides.
\end{abstract}
\maketitle

Negative index matamaterials (NIM) are artificial materials which
have simultaneously negative permittivity and negative
permeability.\cite{VGV,RAS,GVE,AA} In NIM waveguide, modes are
backward when more energy flows in NIM than in other channels.
Coupling of a forward propagating mode in a conventional dielectric
waveguide with a backward mode in a NIM waveguide has been
investigated theoretically in both linear and nonlinear
regimes.\cite{AA,NML} When a forward mode is coupled to a backward
one, the backward mode transports energy backwards, leading to the
formation of spectral gaps without periodic structures along the
waveguide. This feedback mechanism may play an important role in
nanophotonics, as it could significantly simplify complex geometries
that are required for subwavelength optical manipulation and
concentration. However, due to the fabrication complexity and high
losses of NIM, coupling involving NIM is currently not
experimentally feasible  and therefore the mechanism has not
attracted significant attention.

\begin{figure}[htb]
\centerline{\includegraphics[width=8.0cm]{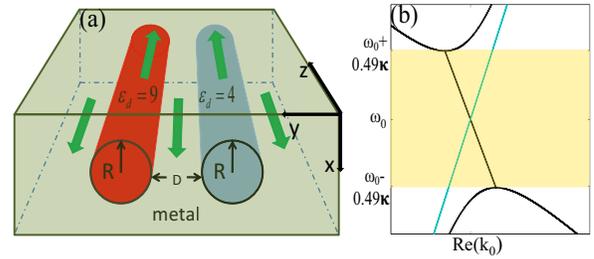}} \caption {(Color
online) (a) Two dielectric waveduides with $\varepsilon_{1}=9$ and
$\varepsilon_{2}=4$ separated by D embedded into infinite metal.
Green arrows indicate the energy flow at different channels for the
wavevector along z; (b) dispersion of two coupled waveguides. Yellow
region indicates the incomplete polarization dependent spectral gap
obtained  using temporal coupled-mode theory with $v_{g}$=0.13c,
$v_{g3}$=0.039c, and $\delta$=0.} \label{fig1}
\end{figure}

There has been a surging interest in the field of plasmonics, as it
offers one of the most promising approaches for subwavelength
optical concentration and manipulation (for a comprehensive review,
see e.g. Refs.\cite{WLB,SAM,DKG,JAS}). In some plasmonic structures,
backward modes exist in regimes when more energy flows in the metal
than in the dielectric.\cite{CAP,JJB,BP,LN,MW}. These structures are
much simpler and more fabricable than those involving NIM. In this
letter, we propose a design of  plasmonic coupler involving the
coupling between the backward TM$_{01}$ and the forward HE$_{11}$
modes in dielectric waveguides embedded into metal [see
Fig.~\ref{fig1}(a)]. We find a polarization dependent spectral gap
in a structure of two coupled waveguides and a complete polarization
independent gap in a C$_{3v}$ structure with four coupled
waveguides.

It was recently reported\cite{YGM} that taking experimental data of
bulk metal\cite{PBJ} in numerical calculations of plasmonic modes
may lead to losses which are much higher than real losses observed
in experiments. In our study, we use the Drude model to simulate the
optical properties of a metal: ${\varepsilon _m}(\omega ) = 1 -
\omega_p^2/\omega({\omega } + i {\omega _\tau })$, where
${\omega_p}$ is the plasma frequency and ${\omega _\tau }$ is the
collision frequency. At the same time, we define two normalized
quantities: loss $\gamma = {\omega _\tau }/{\omega _p}$, and size
parameter $\alpha = R{\omega _p}/c$, where R is the radius of the
dielectric core, and c is speed of light.

\begin{figure}[htb]
\centerline{\includegraphics[width=9cm]{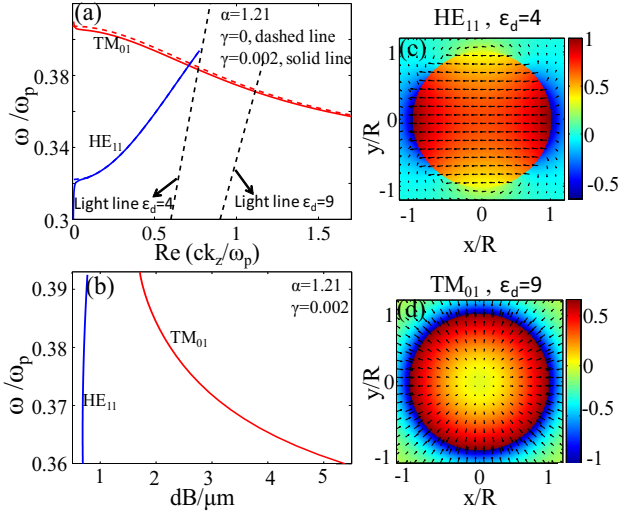}} \caption{(Color
online) (a) Dispersion curve and (b) losses of TM$_{01}$ mode  for
$\varepsilon_d=9$ and HE$_{11}$ mode for $\varepsilon_d=4$. (c) and
(d) Poynting vector component $S_{z}$ (colourmap) and transverse
electric field $E_{t}$ (arrows) for HE$_{11}$ mode and TM$_{01}$
respectively at $\omega/\omega_{p}$=0.3856 with $\gamma$=0.002.}
\label{fig2}
\end{figure}

Fig.~\ref{fig1}(a) shows the two-waveguide structure we study: two
dielectric rods of the same radius $\alpha=1.21$ (corresponding R is
about 25nm for silver) but with $\varepsilon_{1}=9$ and
$\varepsilon_{2}=4$ embedded into infinite metal. First, by
analysing the dispersion of a single waveguide, we find that the
backward TM$_{01}$ mode for $\varepsilon=9$ intersects with the
forward HE$_{11}$ mode for $\varepsilon = 4$ at
$\omega/\omega_p=0.3856$ [see Fig.\ref{fig2}(a)]. This point
corresponds to $\lambda\approx400nm$ for silver. For the TM$_{01}$
mode,  more energy flows in the metal than in the dielectric, which
is similar to the backward SPP on metallic wires.\cite{CAP,HK} It
should be emphasized that the directionality of TM$_{01}$ and
HE$_{11}$ modes are radius dependent: the TM$_{01}$ mode can become
forward when the radius increases, and the HE$_{11}$ mode can become
backward when the radius decreases.\cite{BP}  However, the HE$_{11}$
mode has linear polarization inside the dielectric [see
Fig.~\ref{fig2}(c)], which could be excited directly with a normal
incident wave\cite{HS}, whereas the TM$_{01}$ mode has radial
polarization [see Fig.~\ref{fig2}(d)] with much higher losses in the
coupling region [see Fig.~\ref{fig2}(b)].
Prior to numerical study, we use temporal coupled-mode
theory\cite{GPA,CMS} (TCMT) to get a qualitative understanding of
dispersion relation in the lossless case. The eigenmodes of a
coupled system are expressed as a superposition of individual
waveguide modes:
$\textbf{E}=\sum\nolimits_m{{{A}_m}(z,t){\textbf{E}_m}(x,y){e^{i({\omega
_{m0}} + {\kappa _{mm}})t}}}$, where ${\omega _{m0}} = {\omega _m}$
at $k = {k_0}$ and $\kappa_{mm}$ is self coupling coefficient. For
the two coupled waveguides, three modes can couple to one another:
two forward HE$_{11}$ modes of preferred x and y polarizations,
which could be approximately reconstructed by two orthogonal
eigenmodes of circular polarizations: $A_{1,2}(z,t)$, and backward
TM$_{01}$ mode: $A_{3}(z,t)$. The coupled-mode equations in time
domain are:
\begin{eqnarray}\label{eq:1}
&&i\frac{{\partial {A_{1,2}}(z,t)}}{{\partial t}} +i
{v_{g}}\frac{{\partial {A_{1,2}}(z,t)}}{{\partial z}} +\kappa
{A_3}(z,t){e^{i2\delta t}}{\kern 1pt}  = 0{\kern 1pt} {\kern 1pt}
{\kern 1pt} {\kern 1pt}
{\kern 1pt} {\kern 1pt} {\kern 1pt}\nonumber \\
&&i\frac{{\partial {A_3}(z,t)}} {{\partial t}} +i
{v_{g3}}\frac{{\partial {A_3}(z,t)}} {{\partial z}} +\kappa
\sum\limits_{m = 1}^2 {{A_m}(z,t){e^{ - i2\delta t}}}  = 0.\nonumber
\end{eqnarray}
where $\delta$ = $\frac{1}{2}({\kappa _{33}} + {\omega _{30}} -
{\kappa _{11}} - {\omega _{10}})$ = $\frac{1}{2}({\kappa _{33}} +
{\omega _{30}} - {\kappa _{22}} - {\omega _{20}})$ is the
antisymmetry parameter of two waveguides; ${A_{1,2,3}}$ are
normalized envelopes; ${v_{gi}}{\kern 1pt} {\kern 1pt} ({v_g} =
{v_{g1,g2}} > 0,{\kern 1pt} {\kern 1pt} {\kern 1pt} {v_{g3}} < 0)$
are the group velocities at $\omega_{0}=\omega(k_{0})$; ${\kappa
_{12}}$ = ${\kappa _{21}}$= 0 (mode 1 and 2 are orthogonal), and the
other mutual coupling coefficients are identical: ${\kappa _{ij}} =
{\kappa _{ji}}= {\kappa}$ (i=1,2; j=3). In the coupling region we
ignore the dispersion of group velocities and assume that
${v_{g,g3}}$ and $\kappa$ are constants. By introducing the
following variables: ${a_1}(z,t) = {A_1}(z,t){e^{ - i\delta
t}}{\kern 1pt}$, ${\kern 1pt} {\kern 1pt} {a_2}(z,t) =
{A_2}(z,t){e^{-i\delta t}}$, ${\kern 1pt} {\kern 1pt} {a_3}(z,t) =
{A_3}(z,t){e^{i\delta t}}$ and applying the Fourier transformation,
we obtain the propagation constants of three eigenmodes: ${k_{1,2}}
= \left( {\alpha  \pm i\sqrt { - 8{v_g}_3{v_g}{\kappa ^2} - {\beta
^2}} } \right)/2{v_g}{v_g}_3$, $k_{3} = (\omega  + \delta )/{v_g}$
where $\alpha  = {v_g}(\omega  - \delta ) + {v_{g3}}(\omega + \delta
)$ and  $\beta = {v_g}(\omega  - \delta )-{v_{g3}}(\omega  + \delta
)$. When $- 8{v_g}_3{v_g}{\kappa ^2} \geq{\beta ^2}$, ${k_{1,2}}$ is
a conjugate pair, indicating the existence of a spectral gap, while
$k_3$ corresponds to eignemode $\hat{a}(k,\omega ) =
{\hat{a}_1}(k,\omega ) + {\hat{a}_2}(k,\omega )$, where
${\hat{a}_1}(k,\omega )$ and ${\hat{a}_2}(k,\omega )$ denote
orthogonal circularly polarized modes. Thus, $k_3$ corresponds to
linearly polarized $HE_{11}$ mode, which is not coupled to the
$TM_{01}$ mode. This mode makes the gap polarization dependent.
Fig.~\ref{fig1}(b) shows the results obtained using TCMT of
$\delta=0$ when values of $v_{g,g3}$ are taken from
Fig.~\ref{fig2}(a).

\begin{figure}[htb]
\centerline{\includegraphics[width=9.5cm]{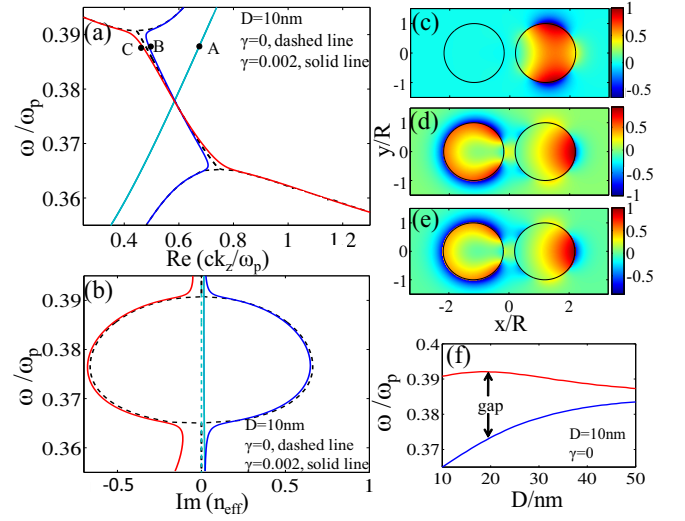}} \caption{(Color
online) (a) Dispersion and (b) losses (imaginary part of ${n_{eff}}
= {k_z}/{k_0} = {k_z}c/\omega $) of the three eignemodes of the
two-waveguide structure. Dashed black ($\gamma$=0), solid red and
green ($\gamma$=0.002) curves correspond to modes of conjugate
propagation constants. Blue curve (dashed curve almost overlaps the
solid  curve as the loss of this mode is comparatively low as shown
in (b)) correspond to the $HE_{11}$ mode that is not coupled to the
$TM_{{01}}$ mode as shown in (c). (c)-(e) $S_{z}$ of modes at the
points (A)-(C) marked in (a), respectively. (f) Gap region vs
distance between waveguides when $\gamma$=0.} \label{fig3}
\end{figure}
Full numerical simulation results using COMSOL (see Fig.~\ref{fig3})
qualitatively agree with TCMT. In the lossless case $\gamma=0$, the
spectral gap is defined by a pair of complex conjugated propagation
constants [see Figs.~\ref{fig3}(a) and (b)]. The gap width increases
with decreasing the distance $D$  [see Fig.~\ref{fig3}(f)], because
the coupling coefficient becomes larger.  When we incorporate some
losses ($\gamma=0.002$), all modes become complex, and the
definition of width of the gap depends on how far it is from the
observing point to the source. However, the gap width of lossless
metal ($\gamma=0$) may still serve as a guide and effective
approximation as shown in Figs.~\ref{fig3}(a) and (b).
\begin{figure}[htb]
\centerline{\includegraphics[width=9cm]{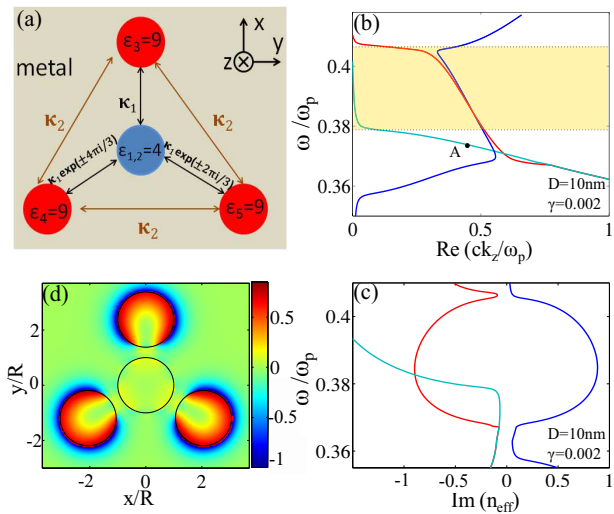}} \caption{(Color
online) (a) Schematic of the four-waveguide structure with $C_{3v}$
symmetry. The distance between $\varepsilon_{1,2} $ waveguide to
$\varepsilon_{3,4,5} $ waveguides is D. (b) Dispersion and (c)
losses of three eignemodes. Blue curve corresponds to the E3 mode.
The complete gap region in lossless case is colored yellow. (d)
$S_{z}$ of E3 mode at point (A) marked in (b).} \label{fig4}
\end{figure}
In addition to the modes of conjugate propagation constants, there
exists one more HE$_{11}$-like decoupled mode. The energy flow of
this mode is mostly confined inside $\varepsilon_{1}=4$ waveguide
[see Fig.~\ref{fig3}(c)]. Thus, the gap of the two coupled
waveguides is incomplete and polarization dependent. To make modes
of different preferred polarization directions degenerate and obtain
a full gap, symmetric structures could be used.\cite{PRM,MJS} One of
the options is to utilize four-waveguide ${C_{3v}}$ structure [see
Fig.~\ref{fig4}(a)]. We use subscripts $n=1,2$ to denote two
$HE_{11}$ modes of circular polarizations and $n=3,4,5$ for three
TM$_{01}$ modes. Based on the symmetry and energy conservation law
in the lossless case, the following relations are satisfied for
mutual coupling coefficients: $\kappa _{12}= \kappa _{12}= 0$,
${\kappa _{1m}}= \kappa^{*} _{m1} = \kappa _{2m}^*= {\kappa
_{m2}}=\kappa _1e^{\frac{2}{3}\pi (m- 3)i}$ for $m,n =3,4,5$ and $m
\ne n$. Due to the $C_{3v}$ symmetry, eigenmodes $\hat{a}(k{\kern
1pt} ,{\kern 1pt} \omega ) = \sum\nolimits_{m = 1}^5 {{\beta
_m}{\hat{a}_m}(k{\kern 1pt} ,{\kern 1pt} \omega )}$ of preferred $x$
polarization $({\beta _1} = {\beta _2})$ and those of preferred $y$
polarization (${\beta _1}$ =${-\beta _2})$ should be
degenerate.\cite{PRM,MJS}  Thus using TCMT we can find five
eigenmodes of three frequencies: ${k_{1,2}} = \left( {\alpha \pm
i\sqrt {- 12{v_g}_3{v_g}{\kappa ^2} - {\beta ^2}} }
\right)/2{v_g}{v_g}_3{\kern 1pt} {\kern 1pt}$ (corresponding to two
degenerate pairs of modes), $\omega = {\omega _3}(k) + 2{\kappa
_2}$, where $\alpha = {v_g}(\omega  - \delta + {\kappa _2}) +
{v_{g3}}(\omega + \delta ),{\kern 1pt} {\kern 1pt} \beta  =
{v_g}(\omega  - \delta + {\kappa _2}) - {v_{g3}}(\omega  + \delta )$
and  $\omega_{3}(k)$ is the dispersion of individual $TM_{01}$ mode.
Again ${k_{1,2}}$ can be a conjugate pair, indicating the existence
of a gap. $\omega = {\omega _3}(k) + 2{\kappa _2}$ corresponds to
eignemode (E3 mode) $\hat a(k,\omega ) = \sum\nolimits_{m = 3}^5
{{{\hat a}_m}(k,\omega )}$, which is a symmetric combination of
$TM_{01}$ modes. The cut off frequency of E3 mode is shifted by
2$\kappa_{2}$ compared with individual $TM_{01}$ mode. Numerical
results from COMSOL are shown in Fig.~\ref{fig4}. This allows us to
conclude that the spectral gap indicated by the yellow region of
four coupled waveguides becomes polarization independent when E3
mode is cutoff. For larger losses (metal in deep ultraviolet regime)
the spectral gap still exists, but the effective width becomes
smaller and eventually disappears as increasing losses make the
differences between gap and non-gap region smaller. To enable
coupling at longer-wavelength regime, where losses of metal is
lower, one could use dielectric waveguides with higher
permittivities (GaAs for example).

In summary, we have studied a coupler based on two dielectric
waveguides in metal involving the coupling of backward and forward
waves. By using the temporal coupled-mode theory we have predicted a
spectral gap in such a system without a periodic structure. This
result has been verified by direct numerical simulations. Moreover,
we have demonstrated that a complete polarization independent gap
can be achieved by using four coupled waveguides with $C_{3v}$
symmetry. Similar coupling between surface plasmon polaritons (SPPs)
can happen in metallic-wire structures when the radius is small
enough to support backward SPPs.\cite{CAP} However, high losses of
backward SPPs on metallic wires prevent them from realistic
realizations. We anticipate that by incorporating materials with
gain and/or nonlineararities, the proposed structure can be
considered as a new platform for the study of gap solitons, optical
bistability, high-Q cavities, plasmonic nanolaser in various systems
without periodicity.

The authors acknowledge a financial support from the Australian
Research Council and useful discussions with B. T. Kuhlmey, I. V.
Shadrivov, A. R. Davoyan, T. P. White,  D. A. Powell, R. Iliew, A.
S. Solntsev, and J. F. Zhang.


\begin{thebibliography}{99}
\bibitem{VGV} V. G. Veselago, Sov. Phys. Usp. \textbf{10}, 509 (1968).
\bibitem{RAS} R. A. Shelby, D. R. Smith,
and S. Schultz, Science 292, \textbf{77}(2001).
\bibitem{GVE} G. V. Eleftheriades, and  K. G. Balmain, {\em Negative
Refraction Metamaterials: Fundamental Principles and Applications}
(Wiley, New Jersy, 2005).
\bibitem{AA}  A. Alu and N. Engheta, in {\em Negative-Refraction
Metamaterials}, edited by G.V. Eleftheriades and K.G. Balmain
(Wiley, New York, 2005).
\bibitem{NML} N. M. Litchinitser, I. R. Gabitov, and A. I. Maimistov, Phys. Rev. Lett. \textbf{99}, 113902 (2007).
\bibitem{WLB} W. L. Barnes, A. Dereux, and T. W. Ebbesen, Nature \textbf{424}, 824 (2003).
\bibitem{SAM} S. A. Maier, {\em Plasmonics: Fundamentals and Applications} (Springer-Verlag, Berlin, 2007).
\bibitem{DKG} D. K. Gramotnev and S. I. Bozhevolnyi, Nat. Photon. \textbf{4}, 83 (2010).
\bibitem{JAS} J. A. Schuller, E. S. Barnard, W. Cai, Y. C. Jun, J. S.
White, and M. L. Brongersma, Nat. Mater. \textbf{9}, 193 (2010).
\bibitem{CAP} C. A. Pfeiffer, E. N. Economou, and K. L. Ngai, Phys. Rev. B \textbf{10}, 3038
(1974).
\bibitem{JJB} J. J. Burke, G. I. Stegeman, and T. Tamir,  Phys. Rev. B \textbf{33}, 5186 (1986).
\bibitem{BP}  B. Prade and J. Y. Vinet, J. Lightwave Technol. \textbf{12}, 6 (1994).
\bibitem{LN}  L. Novotny and C. Hafner, Phys. Rev. E \textbf{50}, 4094 (1994).
\bibitem{MW}  M. Wegener, G. Dolling, and S. Linden, Nat. Mater. \textbf{6}, 475 (2007).
\bibitem{YGM} Y. Ma, X. Li, H. Yu, L. Tong, Y. Gu, and
Q. Gong, Opt. Lett. \textbf{35}, 1160 (2010).
\bibitem{PBJ} P. B. Johnson and R. W. Christy, Phys. Rev. B \textbf{6}, 4370 (1972).
\bibitem{HK}  H. Khosravi, D. R. Tilley, and R. Loudon, J. Opt. Soc. Am. A \textbf{8}, 112 (1991).
\bibitem{HS}  H. Shin, P. B. Catrysse, and S. Fan, Phys. Rev. B \textbf{72}, 085436 (2005).
\bibitem{GPA} G. P. Agrawal, {\em Nonlinear Fiber Optics} (Elsevier, San Diego, 2007).
\bibitem{CMS} C. Martijn de. Sterke, D. G. Salinas, and J. E. Sipe, Phys. Rev. E  \textbf{54}, 1969
(1996).
\bibitem{PRM} P. R. McIsaac, IEEE Trans. Microwave Theory Tech. \textbf{23}, 421 (1975).
\bibitem{MJS} M. J. Steel, T. P. White,  C. Martijn de. Sterke, R. C. McPhedran, and
L. C. Botten, Opt. Lett., \textbf{26}, 488 (2001).
\end{thebibliography}
\end{document}